\newlength{\absize}
\documentclass[12pt]{article}

\usepackage{psfig}
\usepackage{epsfig}
\usepackage{amssymb}
\usepackage{hyperref}

\renewcommand{\baselinestretch}{1.5}

\begin{document}
\thispagestyle{empty}
\pagestyle{empty}
\renewcommand{\thefootnote}{\fnsymbol{footnote}}
\newcommand{\starttext}{\newpage\normalsize
\pagestyle{plain}
\setlength{\baselineskip}{3ex}\par
\setcounter{footnote}{0}
\renewcommand{\thefootnote}{\arabic{footnote}}
}

\newcommand{\preprint}[1]{\begin{flushright}
\setlength{\baselineskip}{3ex}#1\end{flushright}}
\renewcommand{\title}[1]{\begin{center}\LARGE
#1\end{center}\par}
\renewcommand{\author}[1]{\vspace{2ex}{\Large\begin{center}
\setlength{\baselineskip}{3ex}#1\par\end{center}}}
\renewcommand{\thanks}[1]{\footnote{#1}}
\renewcommand{\abstract}[1]{\vspace{2ex}\normalsize\begin{center}
\centerline{\bf Abstract}\par\vspace{2ex}\parbox{\absize}{#1
\setlength{\baselineskip}{2.5ex}\par}
\end{center}}

\newlength{\eqnparsize}
\setlength{\eqnparsize}{.95\textwidth}

\setlength{\jot}{1.5ex}
\newcommand{\figsize}{\small}
\renewcommand{\bar}{\overline}
\font\fiverm=cmr5
\input prepictex
\input pictex
\input postpictex
\input{psfig.sty}
\newdimen\tdim
\tdim=\unitlength

\setcounter{bottomnumber}{2}
\setcounter{topnumber}{3}
\setcounter{totalnumber}{4}
\renewcommand{\bottomfraction}{1}
\renewcommand{\topfraction}{1}
\renewcommand{\textfraction}{0}

\def\draft{
\renewcommand{\label}[1]{{\quad[\sf ##1]}}
\renewcommand{\ref}[1]{{[\sf ##1]}}
\renewenvironment{thebibliography}{\section*{References}}{}
\renewcommand{\cite}[1]{{\sf[##1]}}
\renewcommand{\bibitem}[1]{\par\noindent{\sf[##1]}}
}

\def\theequation{\thesection.\arabic{equation}}
\preprint{\#HUTP-04/A002}
\title{On Exact Superpotentials, Free Energies and Matrix Models}
\author{Girma Hailu\thanks{hailu@feynman.harvard.edu}
and Howard Georgi\thanks{georgi@physics.harvard.edu}
\\
\small\sl Jefferson Laboratory of Physics\\
\small\sl Harvard University \\
\small\sl Cambridge, MA 02138 }


\abstract{We discuss exact results for the full
nonperturbative effective superpotentials of four dimensional
$\mathcal{N}=1$ supersymmetric $U(N)$ gauge theories with
additional chiral superfield in the adjoint representation and the
free energies of the related zero dimensional bosonic
matrix models with polynomial potentials in the planar
limit using the Dijkgraaf-Vafa matrix model prescription and
integrating in and out. The exact effective superpotentials are produced
including the leading Veneziano-Yankielowicz term directly from the
matrix models. We also discuss how to use integrating in and out as a tool
to do random matrix integrals in the large $N$
limit.}

\starttext


\setcounter{equation}{0}

\section{Introduction\label{intr}}

The exact nonperturbative effective superpotential of
$\mathcal{N}=2$ supersymmetric gauge theories with all instanton
corrections is completely determined with a knowledge of the
perturbative UV physics, the singularities in moduli space and the
monodromies around the singularities. \cite{SW-1}   The field
content of a pure $\mathcal{N}=2$ gauge theory is a vector field
$A_{\mu}$, two Weyl fermions $\lambda_{\alpha}$ and
$\psi_{\alpha}$, and a complex scalar $A$. In $\mathcal{N}=1$
language, these are a combination of a field strength chiral
superfield $W_{\alpha}$ containing $\lambda_{\alpha}$ and
$A_{\mu}$, and a scalar chiral superfield $\Phi$ containing $A$
and $\psi_{\alpha}$ all transforming in the adjoint representation. Suppose
this pure $\mathcal{N}=2$ theory is perturbed by a tree level
superpotential including a mass term for $\Phi$. When one
integrates out $\Phi$, the low energy theory below the mass of
$\Phi$ reduces to $\mathcal{N}=1$. About a decade ago, various
examples of exact superpotentials of $\mathcal{N}=1$
supersymmetric gauge theories were obtained making use of
holomorphy and symmetry arguments. See \cite{IS-R} for a review.
However, for such general tree level perturbations, combinations
of parameters which are not protected by symmetries appear. These
parameters can come in the effective superpotential with any
degree. Dijkgraaf and Vafa \cite{DV-1} found that the effective
superpotentials of $\mathcal{N}=1$ theories obtained by such deformations
of $\mathcal{N}=2$ could be computed by using matrix models.

Consider pure $\mathcal{N}=1$ supersymmetric $U(N)$ gauge theory.
The $SU(N)$ subgroup confines in the infrared and using the one loop
running of the gauge coupling coefficient the Lagrangian that describes
the theory at the cut off can be written as
\begin{equation}
\mathcal{L}=\int d^{2}\theta\, 3N
\log(\frac{\Lambda_{1}}{\Lambda_{1c}})S+ c.c.\, ,\label{eq:rev4-3}
\end{equation}
where $S$ is the glueball superfield defined in terms of the gauge
chiral superfield as
\begin{equation}
S=-\frac{1}{32\pi^2}\mathrm{Tr\,}W^{\alpha}W_{\alpha}\, ,\label{eq:rev4-2}
\end{equation}
$\Lambda_{1}$ is the scale of the $\mathcal{N}=1$ $SU(N)$ gauge
theory and $\Lambda_{1c}$ is the UV cutoff. The point is that from
(\ref{eq:rev4-3}) one sees that $3N\log(\Lambda_{1})$ linearly
couples to $S$ and acts as its source. On the other hand, gaugino
condensation gives the nonperturbative effective superpotential
\begin{equation}
W_{\mathrm{eff}}=N\Lambda_{1}^{3}.\label{eq:rev4-4}
\end{equation}
Note that we have suppressed
the $N$ phases $e^{2\pi i k/N}$, $k=0$ to $N-1$, in (\ref{eq:rev4-4})
due to the breaking of the $Z_{2N}$ $R$ symmetry down to $Z_{2}$
by gaugino condensation. We can integrate in $S$ to
(\ref{eq:rev4-4}) with $3N\log(\Lambda_{1})$ as its source and
calculate the glueball superpotential while still keeping
$\Lambda_{1}$ as a parameter in the theory by introducing an
auxiliary field $A$ and minimizing
\begin{equation}
W=NA^3-3NS \log(\frac{A}{\Lambda_{1}})\label{wintins}
\end{equation}
with $A$ which gives the Veneziano-Yankielowicz effective
superpotential \cite{VY}
\begin{equation}
W_{\mathrm{eff}}=NS-NS\log(\frac{S}{\Lambda_{1}^{3}}).\label{eq:rev4-5}
\end{equation}
Integrating out $S$ in (\ref{wintins}) simply gives back the
original nonperturbative superpotential (\ref{eq:rev4-4}).

Suppose one adds a chiral superfield $\Phi$ in the adjoint
representation with a tree level superpotential
\begin{equation}
W_{\mathrm{tree}}=\sum_{p=1}^{n}\frac{g_{p+1}}{p+1}\mathrm{Tr\,}\Phi^{p+1}.\label{wtree}
\end{equation}
to the $\mathcal{N}=1$ $U(N)$ gauge theory. For instance, for the
specific case of a cubic tree level superpotential with $n=2$ in
(\ref{wtree}) and the gauge symmetry in the low energy theory
unbroken, there is a combination of parameters $g_{3}^2S/g_{2}^3$
that has no charge under all symmetries. This parameter can appear
with any power and the instanton corrections to the effective
nonperturbative superpotential can be written as
$\sum_{k=1}^{\infty}c_{k}(g_{3}^2S/g_{2}^3)^{k} S$ where $c_{k}$
are constants. Our interest is to find exact analytic expressions
for the glueball effective superpotential whose series expansion
produces all terms including the leading Veneziano-Yankielowicz
and the instanton corrections and to compute the free energies of
the corresponding zero dimensional bosonic matrix models using
integrating in \cite{ILS-1} and out techniques. The relation between
the Dijkgraaf-Vafa matrix model prescription and the
integrating in method for $U(N)$ gauge theories was discussed in
\cite{F1, CIV} and our approach on the integrating in side here
will be along similar lines.

For completeness we will start with a very brief review of a zero
dimensional $U(N)$ bosonic matrix model as originally developed in
\cite{BIPZ} and reviewed in \cite{DGZ}. We will then compute the
complete exact nonperturbative superpotentials for a quadratic,
cubic and quartic tree level superpotentials using the
Dijkgraaf-Vafa matrix model prescription. The free energies in the
matrix model for cubic and quartic potentials were originally
computed in \cite{BIPZ} and the corresponding instanton
corrections to effective superpotentials were computed in
\cite{FO, AGH} using the Dijkgraaf-Vafa prescription. However, the
leading Veneziano-Yankielowicz term was added by hand. Here we
will point out the proper normalization of the partition function
and scheme that produces the exact effective superpotential
including the Veneziano-Yankielowicz term directly from the matrix
model. We will then continue with computing the full exact
effective superpotential for quadratic, cubic and quartic tree
level deformations using integrating in and out. The results from
the two approaches exactly agree to all order including the
leading Veneziano-Yankielowicz term and the instanton corrections.
We will also provide a scheme for using integrating in and out to
do random matrix integrals. For the case of one-cut $U(N)$ bosonic
matrix models we will discuss in this note, integrating in and out
combined with the Dijkgraaf-Vafa matrix model prescription
provides a very simple tool to do random matrix integrals in the
planar limit for any tree level polynomial potential.

\setcounter{equation}{0}

\section{Matrix model\label{matrint}}

Consider a zero dimensional $U(N)$ bosonic matrix model with a
potential given by (\ref{wtree}). $\Phi$ is an $N\times N$ matrix.
Extremizing the potential generically gives $n$ distinct values
$a_{I}$ of $\Phi$, where $I=1$ to $n$. If $\Phi$ is taken to have
classical eigenvalues $a_{I}$ each with degeneracy $N_{I}$ such
that $N=\sum_{I=1}^{n}N_{I}$, then the gauge symmetry in the low
energy theory is broken to $\prod_{I=1}^{n}{U(N_{I})}$ . Our
interest in this note is the case where all $a_{I}$ are equal and
the gauge symmetry is preserved. We will take $a_{I}=0$.
Now consider the partition function
\begin{equation}
Z=\frac{1}{\mathrm{Vol}(U(N))}\int d\Phi
\,e^{-\frac{1}{g_{s}}W_{\mathrm{tree}}}.\label{mi}
\end{equation}
The large $N$ (or small $g_{s}$) limit is done with  $N\rightarrow
\infty$ and $g_{s}\rightarrow 0$ where
\begin{equation}
S\equiv Ng_{s}\label{s}
\end{equation}
is fixed. The perturbative expansion of (\ref{mi}) leads to a sum
of Feynman diagrams over Riemann surfaces. In the large $N$ limit
only the leading order sum over a genus zero planar surface
survives and the partition function becomes
\begin{equation}
Z=e^{-\frac{\mathcal{F}_{0}}{g_{s}^{2}}},\label{z0}
\end{equation}
where $\mathcal{F}_{0}$ is the free energy in the planar limit.

In the large $N$ limit, the integral is done using the standard
matrix model technology where one starts with transforming the
integral from $\Phi$ to a set of eigenvalues $\lambda_{i}$ of
$\Phi$. The matrix integral (\ref{mi}) when transformed to integral over eigenvalues
$\lambda_{i}$ becomes
\begin{equation}
Z=\int\prod_{i}d\lambda_{i}\,e^{-\frac{1}{g_{s}}\sum_{i}W(\lambda_{i})
+2\sum_{i<j}\mathrm{log}|\lambda_{i}-\lambda_{j}|},\label{mi2}
\end{equation}
where we write the tree level superpotential in (\ref{wtree}) as
$W_{\mathrm{tree}}(\Phi)=\mathrm{Tr}\,W(\Phi)$. The indices $i$ and $j$
in this section run over $1$ to $N$. The second term in
the exponent of the integrand in (\ref{mi2}) comes from the
Jacobian in the transformation of the measure of the integral from
the independent components of $\Phi$ to the eigenvalues. This term
gives rise to a repulsive interaction between the eigenvalues
resulting in distributions of eigenvalues around each extremum
point. In the case where the original gauge symmetry is preserved
in the low energy theory and the extremum point is taken to be at the origin,
all the eigenvalues are distributed
around the origin. The equation of motion of the eigenvalues is
\begin{equation}
\frac{1}{g_{s}}W'(\lambda_{i})-2\sum_{j\ne i}\frac{1}{\lambda_{i}-\lambda_{j}}=0.\label{eig}
\end{equation}
Equation (\ref{eig}) is difficult to solve directly in practice
and one introduces the resolvent
\begin{equation}
w(x)\equiv\frac{1}{N}\sum_{i}\frac{1}{\lambda_{i}-x},\label{w}
\end{equation}
where $x$ is complex. Note that $w(x)$ has the large $x$
asymptotic behavior
\begin{equation}
w(x)\to -1/x.\label{wasy}
\end{equation}
The equation of motion of the eigenvalues in the large $N$ limit
gives a quadratic equation for the resolvent,
\begin{equation}
w(x)^{2}+\frac{1}{S}W'(x)w(x)+\frac{1}{4S^{2}}f(x)=0,\label{wa}
\end{equation}
where
\begin{equation}
f(x)=\frac{4S}{N}\sum_{i}\frac{W'(x)-W'(\lambda_{i})}{x-\lambda_{i}}\label{f}
\end{equation}
is a polynomial function of degree $n-1$. (\ref{wa}) has the
solution
\begin{equation}
w(x)=-\frac{1}{2S}\Bigl(W'(x)\pm\sqrt{W'(x)^{2}-f(x)}\Bigr).\label{wb}
\end{equation}
Because $W'(x)$ and $f(x)$ are polynomial functions, any singular
behavior of $w(x)$ comes from the square root in the second term
in (\ref{wb}). Note that $W'(x)^{2}-f(x)$ is a polynomial of
degree $2n$ and thus $w(x)$ has in general the same number of
branch points. This singularity structure and the asymptotic large
$x$ behavior of $w(x)$ are enough to completely determine $w(x)$
in the one branch cut case where $n-1$ roots of $W'(x)^{2}-f(x)$
come in pair. In the large $N$ limit, one defines the eigenvalue
density,
\begin{equation}
\rho(\lambda)=\frac{1}{N}\sum_{i}\delta(\lambda_{i}-x),\label{rho}
\end{equation}
 such that $\int\rho(\lambda)d\lambda=1$. The point is that
$w(x)$ and $\rho(x)$ are related,
\begin{equation}
w(x)=\int\frac{\rho(\lambda)d\lambda}{\lambda-x}.\label{wrho}
\end{equation}
Therefore, once $w(x)$ is obtained, we can invert (\ref{rho}) to
compute the eigenvalue density,
\begin{equation}
\rho(\lambda)=\frac{1}{2\pi i}\Bigl(w(\lambda+i0)-w(\lambda-i0)\Bigr).\label{rhow}
\end{equation}
With $\rho(\lambda)$ in hand, the free energy in the large $N$ limit follows from
(\ref{z0}) and (\ref{mi2}),
\begin{equation}
\mathcal{F}_{0}=S\int d\lambda\rho(\lambda)W(\lambda)-S^{2}\int\int d\lambda
d\mu\,\rho(\lambda)\rho(\mu)\log|\lambda-\mu|.\label{f0}
\end{equation}
Using the equations of motion given by (\ref{eig}) in the large $N$
limit, (\ref{f0}) can be written in a more convenient form,
\begin{equation}
\mathcal{F}_{0}=\frac{1}{2}S\int d\lambda\,\rho(\lambda)W(\lambda)
-S^{2}\int d\lambda\,\rho(\lambda)\log|\lambda|.\label{f0a}
\end{equation}

Following Dijkgraaf-Vafa \cite{DV-1}, the effective superpotential
is given by
\begin{equation}
W_{\mathrm{eff}}=N\frac{\partial\mathcal{F}_{0}}{\partial
S}+NS\log(\Lambda^{2}),\label{weff}
\end{equation}
where $N\log(\Lambda^{2})$ is a source for $S$  and $\Lambda$ is
the nonperturbative scale in the $\mathcal{N}=2$ gauge theory.

The important point and scheme we want to point out is that with
the proper normalization of the partition function as given by
(\ref{mi}) and the planar limit of the free energy that follows
from (\ref{mi2}), the exact effective superpotential is produced
using (\ref{weff}) without a need for adding the
Veneziano-Yankielowicz term by hand. We will explicitly see in
subsection (\ref{quadp}) that the exact Veneziano-Yankielowicz
superpotential follows from a quadratic tree level potential with
this scheme. We will also derive the appropriate free energies in
the planar limit for cubic and quartic tree level superpotentials
and show that the full exact effective superpotentials including
both the Veneziano-Yankielowicz term and all instanton correction
are produced.

\subsection{Quadratic potential\label{quadp}}

Consider a quadratic mass term with $n=1$ and $g_{2}=m$ in
$W_{\mathrm{tree}}$ given by (\ref{wtree}). Now $W(x)=mx^{2}/2$
has only one extremum point and it occurs at the origin. The most
general form of $w(x)$ follows from (\ref{wasy}) and (\ref{wb}),
\begin{equation}
w(x)=-\frac{1}{2S}(mx-m\sqrt{x^{2}-a^{2}}),\label{wq}
\end{equation}
where $a=2\sqrt{S/m}$. The Wigner semicircle density distribution
of the eigenvalues
follows from (\ref{wq}) in (\ref{rhow}),
\begin{equation}
\rho(\lambda)=\frac{m}{2\pi S}\sqrt{a^{2}-\lambda^{2}},\label{rhoq}
\end{equation}
for $\lambda\in[-a,\, a]$. Note that because $W_{\mathrm{tree}}$
is even, the expectation value of $\Phi$ vanishes and $U(N)$
automatically reduces to $SU(N)$. Using (\ref{rhoq}) and
$W(\lambda)=m\lambda^{2}/2$ in (\ref{f0a}), we obtain the free
energy
\begin{equation}
\mathcal{F}_{0}=\frac{3}{4}S^{2}-\frac{1}{2}S^{2}\log(\frac{S}{m}).\label{foq}
\end{equation}

The effective superpotential follows from (\ref{foq}) in (\ref{weff}),
\begin{equation}
W_{\mathrm{eff}}=NS-NS\log(\frac{S}{\Lambda_{1}^{3}}),\label{vy}
\end{equation}
where $\Lambda_{1}=(\Lambda^{2}m)^{1/3}$ is the scale of the lower
energy pure $\mathcal{N}=1$ theory related to the scale $\Lambda$
of the higher energy $\mathcal{N}=2$ theory by threshold matching
of the gauge coupling running at the energy scale $m$. Now
(\ref{vy}) is exactly the Veneziano-Yankielowicz superpotential.
It was discussed in \cite{DV-1, CDSW} that the volume of the
$U(N)$ gauge group gives a contribution like
$\frac{1}{2}S^{2}\log(S)$ to the free energy in the planar limit.
For a recent discussion on this point and on adding the
Veneziano-Yankielowicz term by hand, see \cite{AGH}. What we have
shown here is that with the proper normalization and scheme
discussed earlier in this section, the appropriate free energy and
the exact Veneziano-Yankielowicz superpotential are obtained.
There is no need for adding the Veneziano-Yankielowicz term by
hand in computing the exact effective superpotential for any tree
level superpotential deformation using this scheme.

\subsection{Cubic potential\label{cubicp}}

Next consider the cubic tree level superpotential,
\begin{equation}
W_{\mathrm{tree}}=\frac{1}{2}m\, \mathrm{Tr}\,(\Phi^{2})+\frac{1}{3}
g\,\mathrm{Tr}\,(\Phi^{3}).\label{wtc}
\end{equation}
Analytic expressions for the free energy of this model and the
quartic potential in the next subsection were first obtained in
\cite{BIPZ} and the instanton corrections in the effective
superpotential using the Dijkgraaf-Vafa prescription have been
computed in \cite{FO} with the Veneziano-Yankielowicz term added by hand.
Here we will compute the effective superpotential
including the Veneziano-Yankielowicz term directly. Now there are
two critical points, one at
$x=0$ and the other at $x=-m/g$. For the case where the $U(N)$
gauge symmetry is unbroken, we need to consider only one
critical point and we choose the one at the origin. In the quantum theory,
all eignevalues
are then distributed in an interval $[a,\,b]$
enclosing the extremum point
at $x=0$. The resolvent is then completely determined noting that $W'(x)^{2}-f(x)$
in (\ref{wb}) has one double root and using the asymptotic behavior (\ref{wasy}),
and it is given by
\begin{equation}
w(x)=-\frac{1}{2S}\Bigl(gx^{2}+mx-(gx+\frac{1}{2}
(a+b)g+m)\sqrt{(x-a)(x-b)}\Bigr),\label{wc1}
\end{equation}
where
\begin{equation}
\frac{1}{4}g(a+b)^{2}+\frac{1}{2}m(a+b)+\frac{1}{8}g(b-a)^{2}=0\label{wc1a}
\end{equation}
and
\begin{equation}
(b-a)^{2}[(a+b)g+m]-16S=0.\label{wc1b}
\end{equation}
Combining the two conditions (\ref{wc1a}) and (\ref{wc1b}) and defining
\begin{equation}
\sigma=\frac{g}{2m}(a+b),\label{sig}
\end{equation}
we obtain
\begin{equation}
\sigma(1+\sigma)(1+2\sigma)+\frac{2g^{2}}{m^{3}}S=0\label{siga}
\end{equation}
with solution
\begin{equation}
\sigma=-\frac{1}{2}+\frac{1}{2\sqrt{3}}(A+\frac{1}{A}),\label{sigb}
\end{equation}
where
\begin{equation}
A=\Bigl(\sqrt{432\frac{g^{4}S^{2}}{m^{6}}-1}
-12\sqrt{3}\frac{g^{2}S}{m^{3}}\Bigr)^{1/3}.\label{sia1}
\end{equation}
We have chosen the solution $\sigma$ to the cubic equation (\ref{siga})
such that $a$ and $b$ are real and the eigenvalues are distributed on the
real axis of $x$ for real ${g^{2}S}/{m^{3}}$.
The eigenvalue density follows from (\ref{wc1}) and (\ref{rhow}),
and writing it in terms of $\sigma$ and making a change of variable,
\begin{equation}
\rho(y)=\frac{1}{2\pi S}\Bigl(gy+(1+2\sigma)m\Bigr)\sqrt{y_{0}^{2}-y^{2}},\label{rhoc}
\end{equation}
where $y=\lambda-m\sigma/g$ and $y_{0}=2\sqrt{S/((1+2\sigma)m)}$.
The eigenvalues are distributed in the range $\lambda\in[a,\, b]$ or
equivalently in $y\in[-y_{0},\, y_{0}]$.

The free energy can then be calculated using (\ref{rhoc}) in (\ref{f0a})
which gives
\begin{equation}
\mathcal{F}_{0}=-\frac{1}{2}S^{2}\log(\frac{S}{m(1+2\sigma)})
+S^{2}\frac{24\sigma^{3}+48\sigma^{2}+37\sigma+9}{12(1+\sigma)(1+2\sigma)^{2}}.\label{f0c}
\end{equation}
Using (\ref{f0c}) in (\ref{weff}) and noting
\begin{equation}
\frac{\partial{\sigma}} {\partial{S}} = \frac {\sigma(1+\sigma)(1+2\sigma)}
{6{\sigma}^2+6\sigma+1}\label{dsis}
\end{equation}
from (\ref{siga}), we obtain \begin{equation}
W_{\mathrm{eff}}=\frac{1}{3}NS\Bigl(2+\frac{2}{1+2\sigma}-\frac{1}{1+\sigma}\Bigr)
-NS\log(\frac{S}{(1+2\sigma)\Lambda_{1}^{3}})\,,\label{weffca}
\end{equation}
where again $\Lambda_{1}=(\Lambda^{2}m)^{1/3}$ is the scale of the low
energy $\mathcal{N}=1$ theory. We will make a series expansion of
(\ref{weffca}) in subsection (\ref{cubicd}) and see that it produces the
exact effective superpotential including the Veneziano-Yankielowicz term
and the instanton corrections to all order.

\subsection{Quartic potential\label{quartp}}

Finally, let us consider the quartic superpotential
\begin{equation}
W_{\mathrm{tree}}=\frac{1}{2}m\,
\mathrm{Tr}\,(\Phi^{2})+\frac{1}{4}g\,\mathrm{Tr}\,(\Phi^{4}).\label{qrt1}
\end{equation}
In this case, again for the one-cut case in which the
gauge symmetry is preserved and all eigenvalues are distributed
around the origin, the resolvent is computed noting that
$W'(x)^{2}-f(x)$ in (\ref{wb}) has two double roots
and using the asymptotic behavior (\ref{wasy}),
\begin{equation}
w(x)=-\frac{1}{2S}\Bigl(gx^{3}+mx-(gx^2+\frac{1}{2}ga^2+m)\sqrt{x^2-a^2}\Bigr),\label{wquart}
\end{equation}
where
\begin{equation}
3ga^4+4ma^2-16S=0.\label{aquart}
\end{equation}
The eigenvalue density follows from (\ref{wquart}) in (\ref{rhow})
\begin{equation}
\rho(\lambda)=\frac{1}{2\pi
S}(g\lambda^2+\frac{1}{2}ga^2+m)\sqrt{a^2-\lambda^2},\label{rquart}
\end{equation}
Using (\ref{qrt1}) and (\ref{rquart}) in (\ref{f0a}), we obtain
the free energy
\begin{equation}
\mathcal{F}_{0}=-\frac{1}{2}S^2
\log(\frac{a^2}{4})+\frac{1}{384}(-m^{2}a^{2}+40 m S a^2+144
S^2).\label{f0mat24}
\end{equation}
Noting from (\ref{aquart}) that
\begin{equation}
\frac{\partial a}{\partial S}=\frac{4}{3ga^{3}+2ma},\label{qdas}
\end{equation}
the effective superpotential is obtained using (\ref{f0mat24}) in
(\ref{weff}),
\begin{equation}
W_{\mathrm{eff}}=\frac{3}{4}NS-NS\log(\frac{a^2}{4})-N\frac{(a^2
m-12S)(a^{2}
m-8S)}{72ga^4+48ma^2}+\frac{5ma^2}{48}N,\label{weffq24}
\end{equation}
where
\begin{equation}
a^{2}=\frac{2m}{3g}(\sqrt{1+12\frac{gS}{m^2}}-1).
\label{asqquartic}
\end{equation}
We will show in subsection (\ref{quartd}) that making a series expansion
of (\ref{weffq24}) gives the correct leading order Veneziano-Yankielowicz
term in addition to all instanton corrections.

\setcounter{equation}{0}
\section{Integrating in and out\label{defint} }

The theory without the tree level deformation is simply pure
\emph{$\mathcal{N}=2$ $U(N)$} gauge theory and it has been well
studied. The quantum moduli space of this theory can be
parameterized by the hyperelliptic curve \cite{SW-1, KLTYAF}
\begin{equation}
y^{2}=P_{N}(x)^{2}-4\Lambda^{2N},\label{he1}
\end{equation}
where $y$ and $x$ are complex coordinated such that
\begin{equation}
P_{N}(x)=\mathrm{det}(x1-\Phi)=\prod_{i=1}^{N}(x-x_{i}),\label{pnx}
\end{equation}
$x_i$ being the diagonal elements of $\Phi$, which can be
made diagonal using D-flatness conditions. A gauge invariant
parametrization of the moduli space can be given by coordinates
\begin{equation}
u_{1}=\mathrm{Tr}\,\Phi=\sum_{i=1}^{N} x_{i}\label{trphi}
\end{equation}
and
\begin{equation}
u_{p}=\frac{1}{p}\mathrm{Tr}\,(\Phi+u_{1}/N)^{p}=
\frac{1}{p}\sum_{i=1}^{N}(x_{i}+u_{1}/N)^{p},\,\,
p=2 \,\mathrm{to}\, N. \label{upx}
\end{equation}

The quantum moduli space given by (\ref{he1}) describes a genus
$N-1$ Riemann surface with two types of cycles. One type
($\beta$-cycle) is related to the handles of the surface and the
second type ($\alpha$-cycle) connects different handles. A dual
scalar field is given by integral of a meromorphic differential
over an $\alpha$-cycle. A dual scalar field vanishes and the
associated monopole (or dyon) becomes massless when an
$\alpha$-cycle degenerates. The strong coupling singularities of
the quantum moduli space correspond to regions where monopoles
become massless. Our interest here is the case where the $U(N)$
gauge symmetry is preserved. This corresponds to the case in which
all the $\alpha$-cycles vanish and the genus $N-1$ Riemann surface
degenerates into a sphere. The effective field theory which
describes the low energy excitations near these singularities is
described in terms of the dual scalar and monopole fields and the
exact effective superpotential is given by, following the
nonrenormalization and linearity principle of \cite{ILS-1},
\begin{equation}
W=\sum_{m=1}^{N-1} A_{D,m}(u)E_{m}\tilde{E}_{m}+
\sum_{p=2}^{n+1}g_{p}u_{p},\label{wmon}
\end{equation}
where $E_{m}$ and $\tilde{E}_{m}$ denote the monopole chiral
multiplet which becomes massless at the $m^{\mathrm{th}}$
singularity and are associated with the dual scalar field
$A_{D,m}$. The second term in (\ref{wmon}) is the tree level
deformation. The steps to compute the effective glueball
superpotential are integrating out $E_{m}$, $\tilde{E}_{m}$,
$u_{p}$ and $u_{1}$ and integrating in $S$. The equations of
motion obtained by minimizing (\ref{wmon}) with $E_{m}$,
$\tilde{E}_{m}$ and $u_{p}$ are
\begin{equation}
A_{D,m}(u)=0,\label{adm0}
\end{equation}
\begin{equation}
\sum_{m=1}^{N-1}\frac{\partial A_{D,m}(u)} {\partial
u_{p}} E_{m}\tilde{E}_{m}+g_{p}=0.\label{eevev}
\end{equation}
Putting (\ref{adm0}) in (\ref{wmon}) gives
\begin{equation}
W_{\mathrm{eff}}=(\sum_{p=2}^{n+1}g_{p}u_{p})|_{A_{D,m}(u)=0}.\label{weff23}
\end{equation}
Now (\ref{eevev}) shows the standard confinement of monopoles and
(\ref{adm0}) expresses the vanishing of the dual scalar fields and
the masslessness of the monopoles at the singularities. The
effective superpotential is thus given by the tree level
deformation evaluated with the constraint that all the dual scalar
fields vanish. The glueball effective superpotential is then
obtained by integrating out $u_1$ and integrating in $S$ while at
the same time keeping the nonperturbative scale as a parameter in
the leading Veneziano-Yankielowicz part of the superpotential.
This can be done by introducing an auxiliary field $A$ and
minimizing
\begin{equation}
\sum_{p=2}^{n+1}(g_{p}u_{p})|_{A_{D,m}(u)=0}\,(\mathrm{with}\,
\Lambda \rightarrow A)-2NS\log(\frac{A}{\Lambda}).\label{weffus}
\end{equation}
with $A$ and $u_{1}$.
Note that when all the $\alpha$-cycles simultaneously degenerate,
$P_{N}(x)^{2}-4\Lambda^{2N}$ has two single roots and $N-1$ double
roots and the hyperelliptic curve factorizes. \cite{DS, CIV}
$P_{N}$ is then given in terms of the Chebyshev polynomial of the
first kind $T_{N}$ \cite{DS},
\begin{equation}
P_{N}(x)=2\Lambda^{N}\,T_{N}(\frac{x}{2\Lambda}),
\end{equation}
where $T_{N}(\frac{x}{2\Lambda})=\cos(N \cos^{-1}(x/(2\Lambda))$.
The diagonal elements $x_{i}$ of $\Phi$ are also parameterized by
\begin{equation}
x_{i}=2\Lambda\cos(\frac{i-1/2}{N}\pi)\label{xif}.
\end{equation}

Once the effective glueball superpotential is obtained, the free
energy in the planar limit is then computed via
\begin{equation}
\mathcal{F}_{0}=\frac{1}{N}\int W_{\mathrm{eff}}(S)dS
-\frac{1}{2}S^2\log(\Lambda^2)\label{f0weff}
\end{equation}
with the boundary condition that when there are other than quadratic
terms in the tree level potential the free energy reduces to
(\ref{foq}) at $g_{p}=0$ for all $p\ne 2$.

We will start with computing the exact effective superpotentials
and the free energies for the quadratic, cubic and quartic tree
level deformations. The results exactly agree
with the matrix model calculations
including the leading Veneziano-Yankielowicz term and all
instanton corrections.

\subsection {Quadratic deformation\label{quadd}}

First consider the quadratic deformation with $n=1$ and $g_{2}=m$
in (\ref{wmon}). In this case, using (\ref{upx}) and (\ref{xif})
in (\ref{weff23}) gives
\begin{equation}
W_{\mathrm{eff}}=mN\Lambda^{2}+\frac{m}{2N}u_{1}^{2}.\label{weff2u}
\end{equation}
We then integrate in $S$ and integrate out $u_{1}$ by minimizing
\begin{equation}
W=mNA^{2}+ \frac{m}{2N}u_{1}^{2} -
2NS\log(\frac{A}{\Lambda})\label{weff2us}
\end{equation}
with $A$ and $u_{1}$.  This gives
\begin{equation}
W_{\mathrm{eff}}=NS-NS\log(\frac{S}{\Lambda_{1}^3}),\label{weff2us1}
\end{equation}
where $\Lambda_{1}=(m\Lambda^2)^{1/3}$ is the scale of the
$\mathcal{N}=1$ theory related to the scale $\Lambda$ of the
$\mathcal{N}=2$ theory by threshold matching of the gauge coupling
running at energy scale $m$. (\ref{weff2us1}) is exactly the
Veneziano-Yankielowicz superpotential. Furthermore, putting
(\ref{weff2us1}) in (\ref{f0weff}) gives the same free energy
(\ref{foq}) as that obtained from the matrix model. In other words,
the matrix integral in the planar limit of a zero dimensional
bosonic matrix model is solved purely using integrating in and out
and the Dijkgraaf-Vafa prescription.

\subsection {Cubic deformation\label{cubicd}}

Next let us consider the cubic potential with $n=2$, $g_{2}= m$
and $g_{3}=g$ in (\ref{wmon}). In this case, putting (\ref{upx}) and
(\ref{xif}) in (\ref{weff23}) gives
\begin{equation}
W_{\mathrm{eff}}=(mN+2gu_{1})\Lambda^{2}+\frac{m}{2N}u_{1}^{2}
+\frac{g}{3N^{2}}u_{1}^{3}.\label{weff23u}
\end{equation}
Minimizing
\begin{equation}
W=(mN+2gu_{1})A^{2}+\frac{m}{2N}u_{1}^{2}+\frac{g}{3N^{2}}u_{1}^{3}
-2NS\log(\frac{A}{\Lambda})\label{weff32us}
\end{equation}
with $A$ and $u_{1}$ gives
\begin{equation}
W_{\mathrm{eff}}=NS-NS\log(\frac{S}{\Lambda_{1}^{3}})
+NS\log(1+\frac{2g}{N}u_{1})+\frac{m}{2N}u_{1}^{2}
+\frac{g}{3N^{2}}u_{1}^{3}\,,\label{weff32us1}
\end{equation}
where
\begin{equation}
u_{1}=-\frac{mN}{2g}+\frac{m^{2}N^{2}}{6B}+\frac{B}{2g^{2}},\label{u13}
\end{equation}
\begin{equation}
B=\Bigl(-4g^{5}N^{3}S+\frac{1}{3\sqrt{3}}N^3g^3 m^3 \sqrt{-1+432\frac{g^{4}S^{2}}{m^6}}\Bigr)^{1/3}\label{au}
\end{equation}
and $\Lambda_{1}=(m\Lambda^2)^{1/3}$.
Expanding (\ref{weffca}) and (\ref{weff32us1}) in $g^{2}S/m^{3}$
gives exactly the same result including the leading
Veneziano-Yankielowicz term and the instanton corrections to all
order,
\begin{eqnarray}
W_{\mathrm{eff}} & = &
NS-NS\log(\frac{S}{\Lambda_{1}^{3}})-2NS\frac{g^{2}S}{m^{3}}
-\frac{32}{3}NS(\frac{g^{2}S}{m^{3}})^{2}\nonumber \\
 &  & -\frac{280}{3}NS(\frac{g^{2}S}{m^{3}})^{3}-1024NS
(\frac{g^{2}S}{m^{3}})^{4}-NS\mathcal{O}
(\frac{g^{2}S}{m^{3}})^{5}.\label{weffcb}
\end{eqnarray}
The free energy can also be computed putting (\ref{weff32us1}) or (\ref{weffcb})
in (\ref{f0weff}).

\subsection {Quartic deformation\label{quartd}}

Consider the quartic potential with $g_{2}= m$, $g_{4}=g$ and all
other $g_{p}=0$ in (\ref{wmon}). Using (\ref{upx}) and (\ref{xif})
in (\ref{weff23}),
\begin{equation}
W_{\mathrm{eff}}=mN\Lambda^{2}+\frac{m}{2N}u_{1}^{2}
+\frac{3gN}{2}\Lambda^{4}+\frac{3g}{N}u_{1}^2\Lambda^{2}
+\frac{g}{4N^3}u_{1}^{4}. \label{weff42n}
\end{equation}
Minimizing
\begin{equation}
W=mNA^{2}+\frac{m}{2N}u_{1}^{2}+\frac{3gN}{2}A^{4}+\frac{3g}{N}u_{1}^2A^{2}
+\frac{g}{4N^3}u_{1}^{4}
-2NS\log(\frac{A}{\Lambda})\label{weff42us}
\end{equation}
with $A$ and $u_{1}$ gives
\begin{equation}
A^2=\frac{m}{6g}(\sqrt{1+\frac{12gS}{m^2}}-1), \quad u_{1}=0.
\label{uaquart}
\end{equation}
Putting (\ref{uaquart}) in (\ref{weff42us}), we obtain the effective glueball
superpotential
\begin{equation}
W_{\mathrm{eff}}=
NS(\frac{1}{2}-\frac{m^{2}}{12gS})
-NS\log(\frac{\sqrt{1+12gS/m^{2}}-1}{6\Lambda_{1}^3 g/m^{2}})
+N \frac{m^2}{12g} \sqrt{1+\frac{12gS}{m^2}}
\,,\label{weff42us1}
\end{equation}
where $\Lambda_{1}=(m\Lambda^2)^{1/3}$.
Making a series expansion of $W_{\mathrm{eff}}$ given by
(\ref{weffq24}) and (\ref{weff42us1}) again gives exactly the same
result,
\begin{eqnarray}
W_{\mathrm{eff}}&=&
NS-NS\log(\frac{S}{\Lambda_{1}^{3}})+\frac{3}{2}NS\frac{gS}{m^2}
-\frac{9}{2}NS(\frac{gS}{m^2})^2\nonumber \\
& & +\frac{45}{2}NS(\frac{gS}{m^2})^3
-\frac{567}{4}NS(\frac{gS}{m^2})^4+NS\mathcal{O}((\frac{gS}{m^2})^5).\label{weffq24s}
\end{eqnarray}
Furthermore, putting (\ref{weff42us1})
in (\ref{f0weff}) and using the boundary condition discussed below (\ref{f0weff}),
the free energy matrix integral in the large $N$ limit
is computed using integrating in and out and the result indeed
agrees with (\ref{f0mat24}).

\section{Conclusion\label{concl}}

We have shown that the Dijkgraaf-Vafa matrix model prescription
with the proper normalization of the partition function and the
scheme we presented gives the exact nonperturbative effective
superpotential without a need for adding the
Veneziano-Yankielowicz term by hand. Both the Dijkgraaf-Vafa
matrix model prescription and the integrating in and out procedure
 give the same result for
the full exact nonperturbative effective superpotential including
the leading Veneziano-Yankielowicz term and all instanton
corrections. It would be important to understand the physical
relations between the two quite different structures, one based on
the large $N$ limit of zero dimensional bosonic matrix models and
the other based on the hypothesis of nonrenormalization and
linearity principles. It is interesting to notice that because the
parametrization of the fully degenerate Seiberg-Witten
hyperelliptic curve was known in terms of the Chebyshev
polynomials, the computation of the effective superpotentials on
the integrating in side reduced to a simple substitution of the
parametrization of the diagonal elements of the adjoint scalar
field into the gauge invariant coordinates and integrating in the
glueball superfield. Consequently, one-cut matrix integrals are
easily computed for any polynomial potential using integrating in
and out. It would be interesting to find the polynomials that
factorize the Seiberg-Witten curve for more general cases where
the degeneration of the quantum moduli space is partial. Once
these polynomials are found, doing multi-cut matrix integrals
should not in principle be a difficult task using integrating in
and out. Integrating in and out can thus be used as an alternative
tool to do random matrix integrals.

\section*{Acknowledgements}

This research is supported in part by the National Science
Foundation under grant number NSF-PHY/98-02709.

\end{document}